\documentclass[cits]{PoS}
\title{CDF Results on Heavy Quarks}

\ShortTitle{CDF Heavy Quarks}

\author{\speaker{Jonathan L. Rosner}\thanks{Presented on behalf of the
CDF Collaboration at Fermilab.}\\
        Enrico Fermi Institute \\
        University of Chicago \\
        E-mail: \email{rosner@hep.uchicago.edu}}

\abstract{Some recent CDF results on heavy quarks (mainly $b$) are presented,
including baryon masses and lifetimes, CP violation in charmless baryon decays,
$B_c$ production, $A_{FB}(b \bar b)$ at high $m(b \bar b)$, and excited
$B$ mesons.}

\FullConference{The 15th International Conference on B-Physics at Frontier
Machines at the University of Edinburgh,\\
		14 -18 July, 2014\\
		University of Edinburgh, UK}

\begin{document}
\section{Introduction}

Precise vertex detection has enabled the CDF Detector at Fermilab to study a
number of properties of mesons and baryons containing heavy quarks.  This talk
is a report on some of the most recent CDF results, primarily on bottom ($b$)
quarks.  All results are based on the full Run II (2001-2011) delivered sample
of 12 fb$^{-1}$; recorded totals range from 8.7 to 9.6 fb$^{-1}$.

\section{Baryon masses and lifetimes \cite{Aaltonen:2014wfa}}

A $\mu^+ \mu^-$ trigger ($p_T(\mu)>1.5$ GeV/$c$) selects events with $J/\psi$,
giving a sample unbiased with respect to $b$-hadron decay time, while a
displaced two-track trigger, biased in favor of long decay times, selects events
with $b$ (and charm) decays.  In the dimuon trigger, tracks in the outer $\mu$
chambers of CDF are matched with those in the central tracker.  In the
two-track trigger, drift chamber tracks provide ``roads'' to an inner silicon
vertex detector \cite{Aaltonen:2013uma}; events are selected with flight
distance greater than 200 $\mu$m from the beam.  

The ground-state $J^P=1/2^+$ baryons with one $b$ quark consist of $\Lambda_b
= bud$, $\Sigma_b^{+,0,-} = b(uu,ud,dd)$, $\Xi_b^{0,-} = bs(u,d)$, and
$\Omega_b^- = bss$.  Previously \cite{CDF:2011ac} CDF had reported the
discovery of four $\Sigma_b^{(*)\pm}$ states, all with the light quarks
coupled up to $J_{\rm light} = 1$.  Now CDF presents updated results on
$\Xi_c^{(0,+)}, \Lambda_b, \Xi_b^{(-,0)}$, and $\Omega_b$ masses and lifetimes.

\begin{figure}[h]
\includegraphics[width=0.33\textwidth]{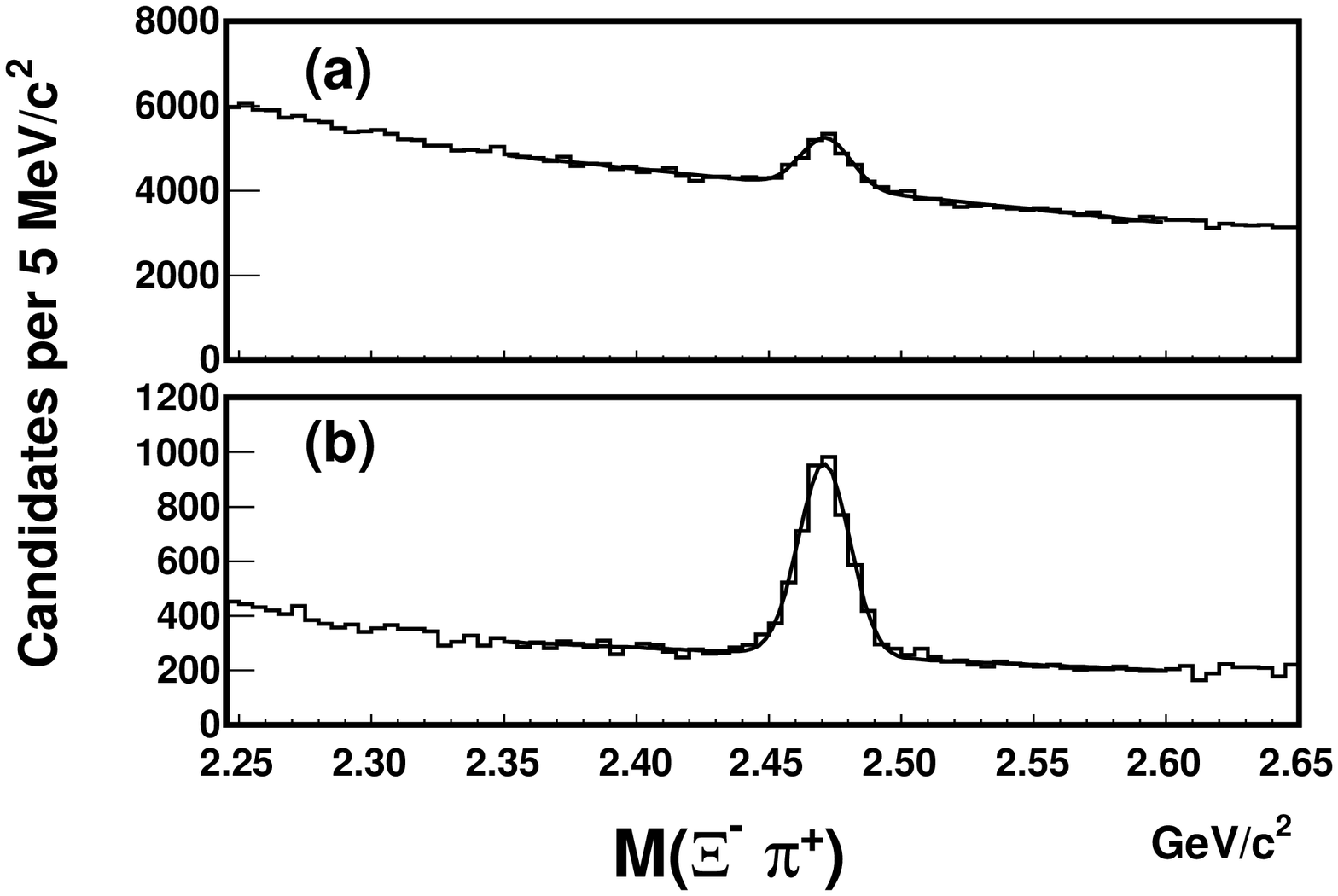}
\includegraphics[width=0.33\textwidth]{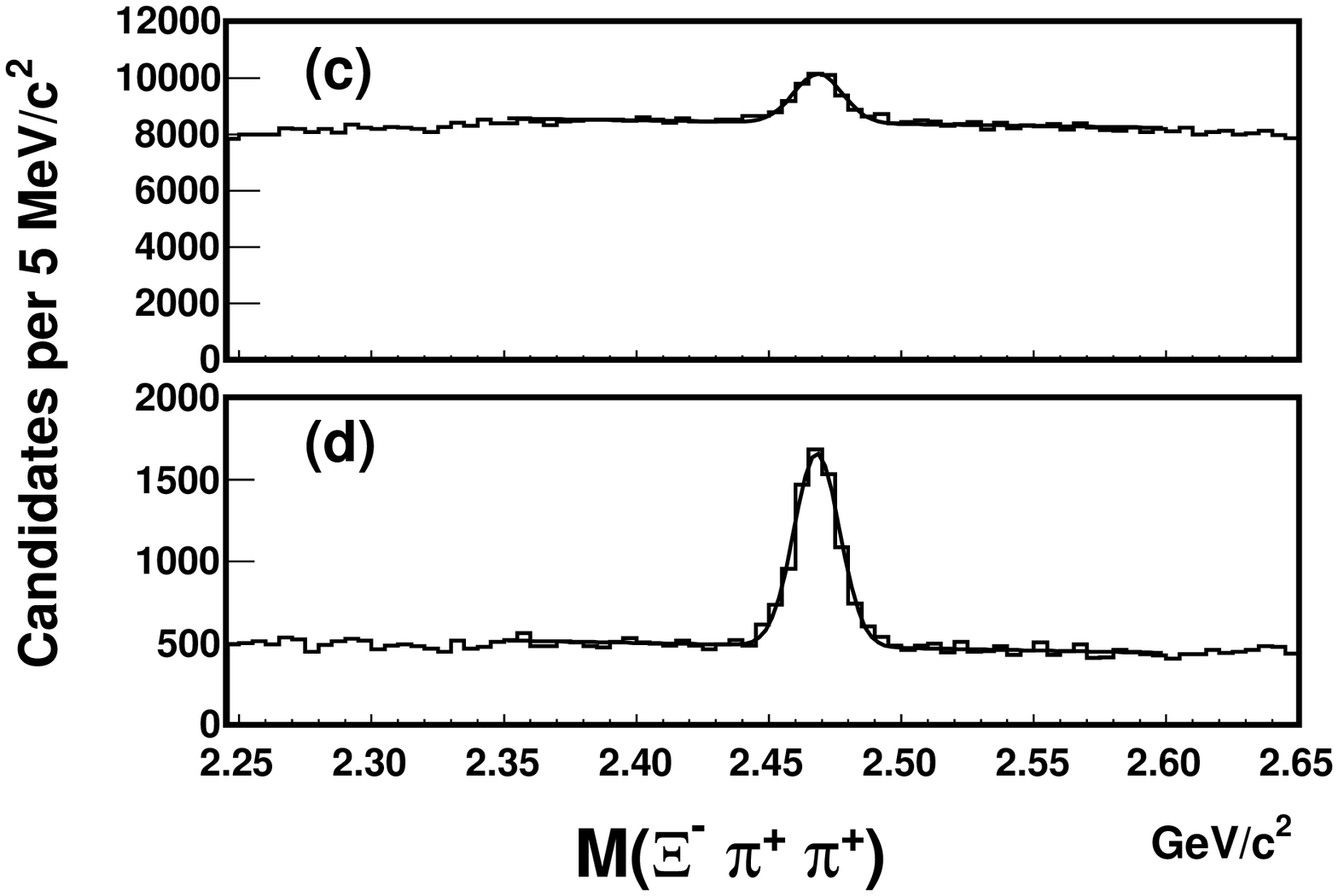}
\includegraphics[width=0.33\textwidth]{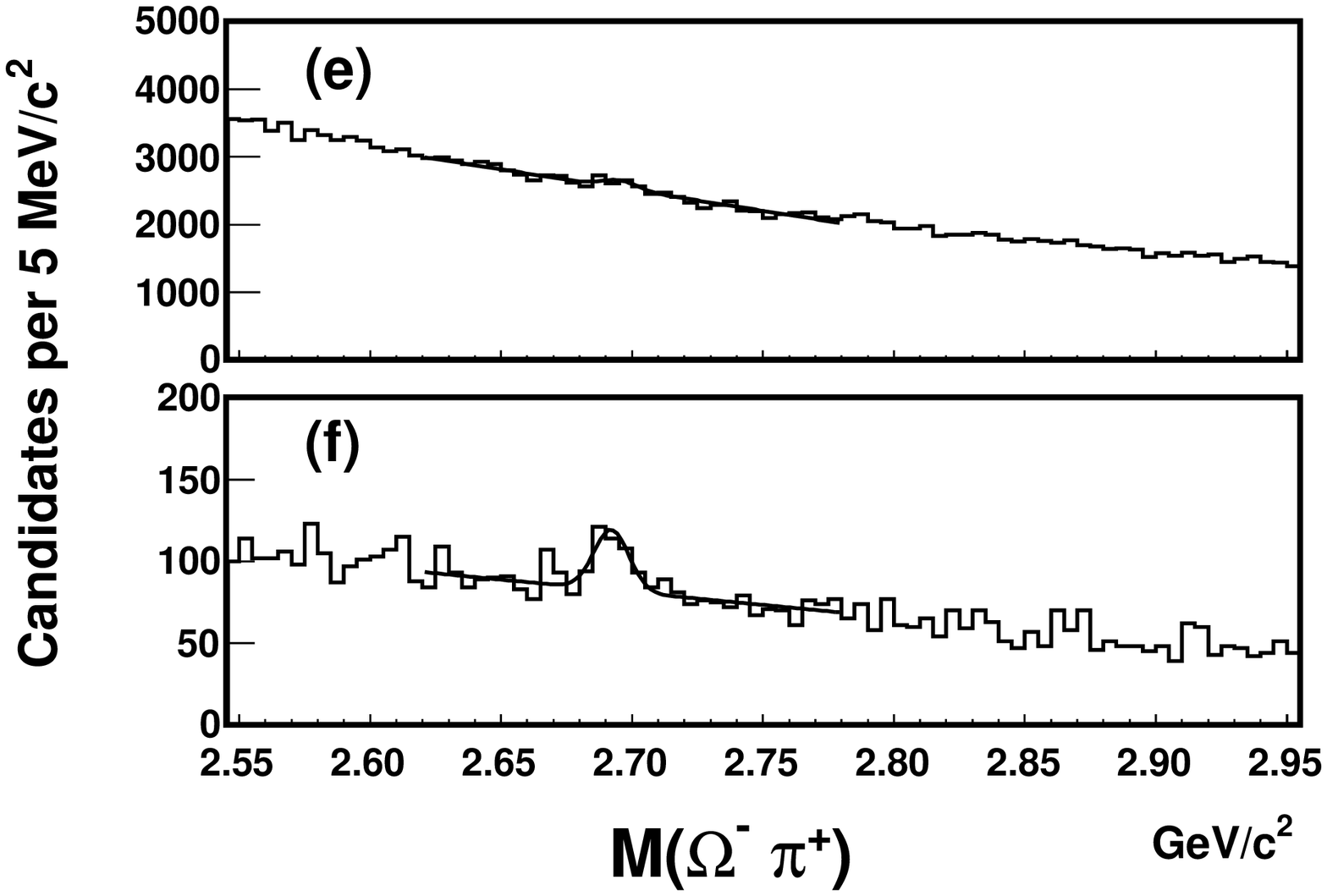}
\caption{(a,b) Signal for $\Xi_c^0 \to \Xi^- \pi^+$ (before, after)
imposition of vertex constraint; (c,d) same for $\Xi_c^0 \to \Xi^-\pi^+ \pi^+$;
(e,f) same for $\Omega_c \to \Omega^- \pi^+$.
\label{fig:charmb}}
\end{figure}

To illustrate the power of good vertex detection, the (top, bottom) plots in
Fig.\ \ref{fig:charmb} denote mass distributions for charmed hyperon
candidates (before, after) the demand that the negative hyperon be tracked in
the silicon detector and the impact parameter with respect to the beam line be
less than 100 $\mu$m.  The signal-to-noise ratio is greatly improved with the
addition of the silicon information, as shown in Table \ref{tab:purity}.

\begin{table}[h]
\caption{Purity of signals for some charmed baryons without and with
information from silicon vertex detector.
\label{tab:purity}}
\begin{center}
\begin{tabular}{c c c} \hline \hline
                 & No silicon & With silicon \\ \hline
$\Xi_c^0$ purity  & $0.15 \pm 0.01$ & $0.63 \pm 0.01$ \\
$\Xi_c^+$ purity  & $0.11 \pm 0.01$ & $0.61 \pm 0.01$ \\
$\Omega_c$ purity & $0.03 \pm 0.01$ & $0.22 \pm 0.05$ \\ \hline
\end{tabular}
\end{center}
\end{table}

The $\Lambda_b$ lifetime has been a long-standing problem.  Theory has favored
$\tau(\Lambda_b)/\tau(B^0)$ close to 1 \cite{JLR,Lenz:2014jha} while many
experiments saw ratios closer to $\sim 0.8$.  With more data and better vertex
detection the observed lifetime has moved closer to theoretical predictions.
The effective mass of $\Lambda_b$ in the $J/\psi \Lambda$ mode has been
plotted in four proper time bins and compared with reference plots for
$B^+ \to J/\psi K^+$ and $B^0 \to (J/\psi K^{*0},J/\psi K_S^0)$ which yield
meson lifetimes within 1\% or better of world averages.

\begin{figure}[h]
\includegraphics[width=0.33\textwidth]{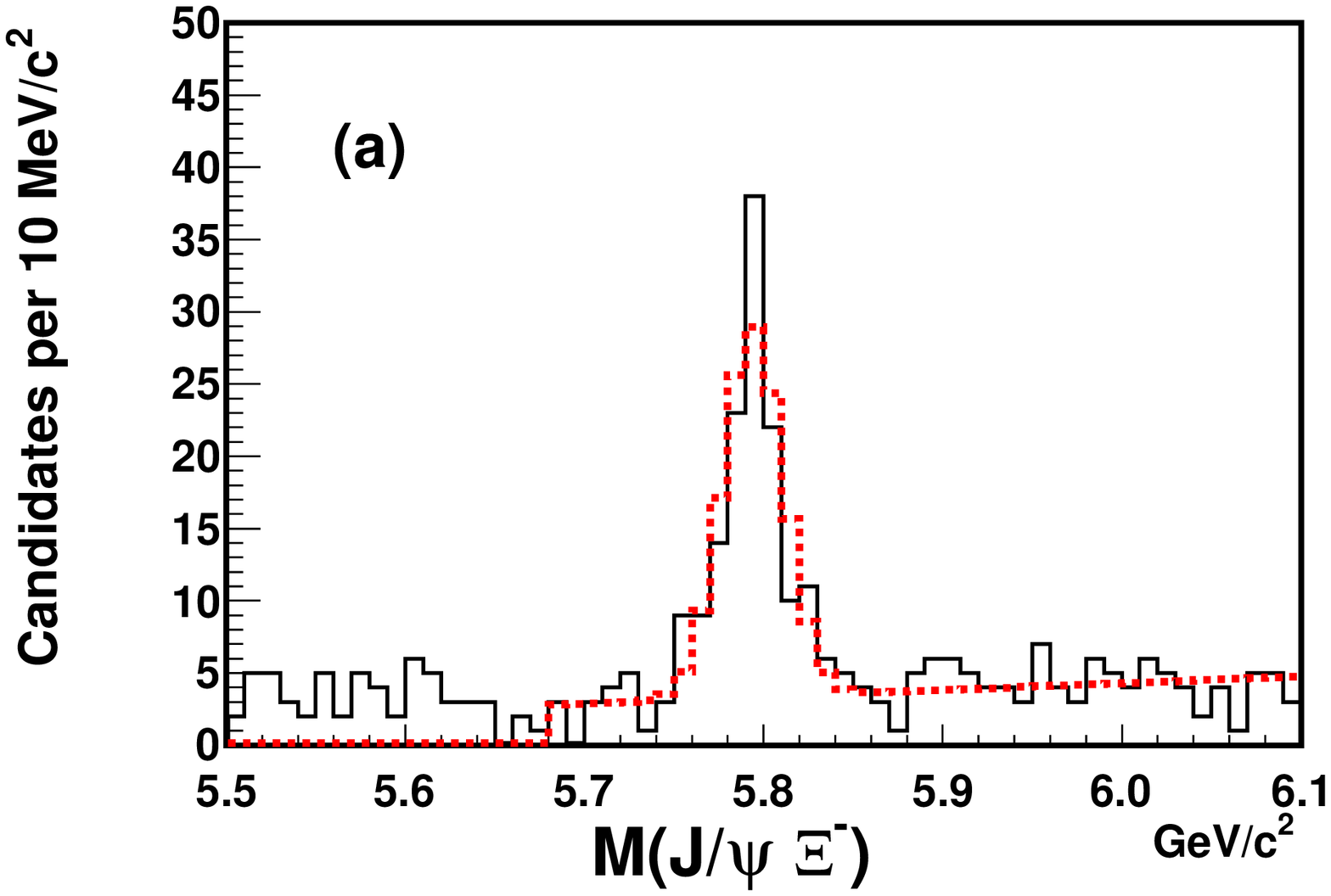}
\includegraphics[width=0.33\textwidth]{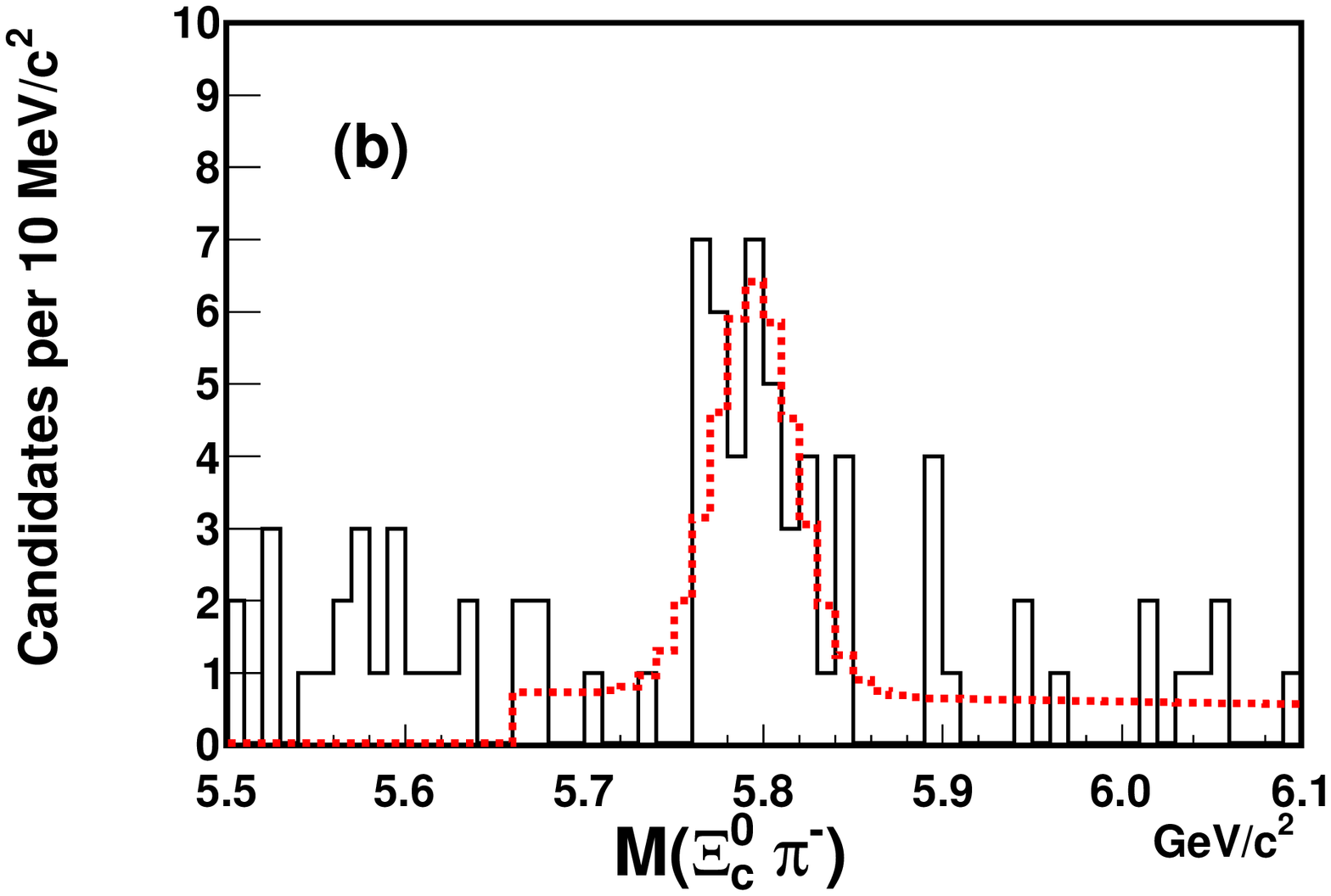}
\includegraphics[width=0.33\textwidth]{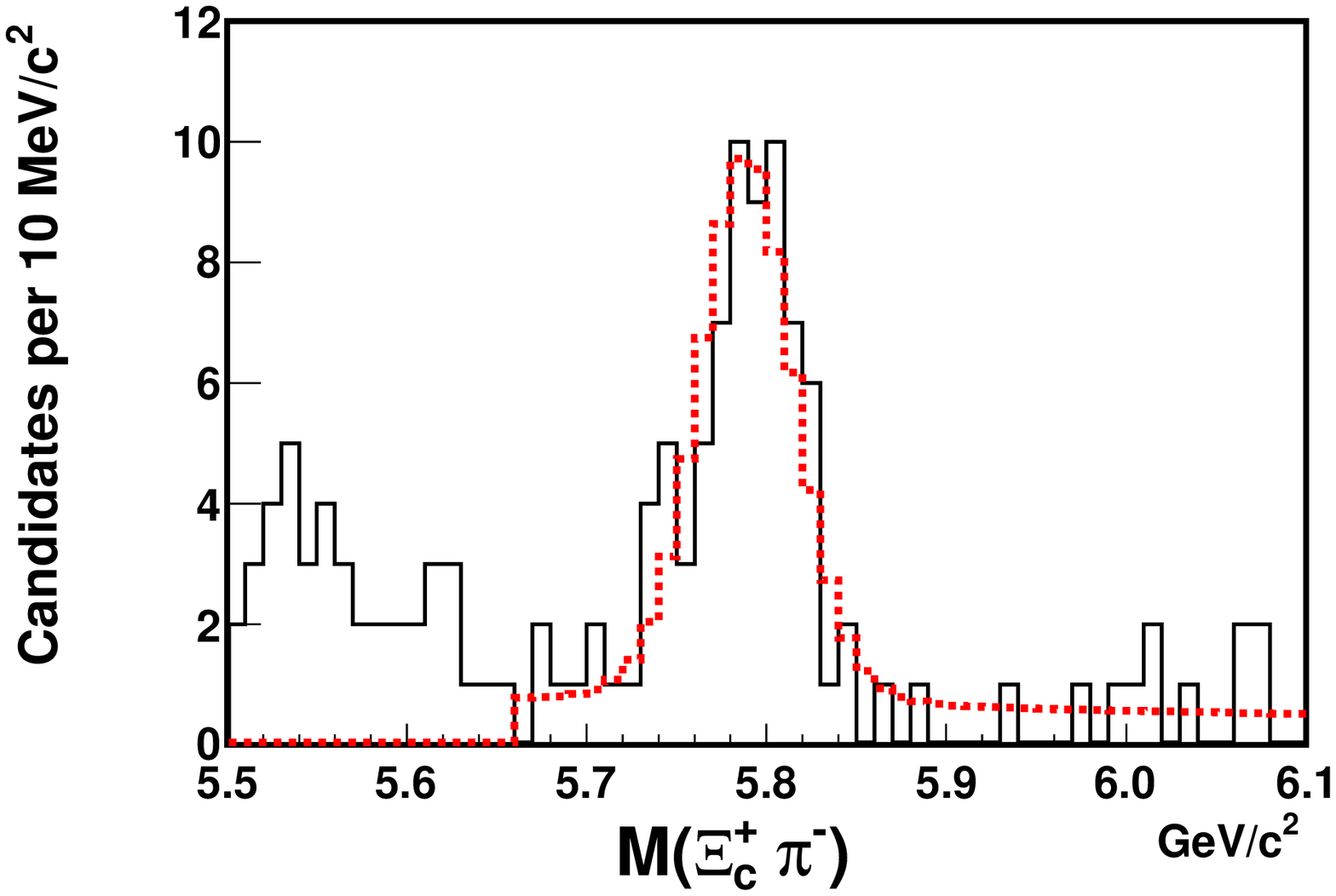}
\caption{Signals for $\Xi_b^- \to J/\psi \Xi^-$ (left); $\Xi_b^- \to \Xi_c^0
\pi^-$ (middle); $\Xi_b^0 \to \Xi_c^+ \pi^-$ (right).  Solid black curves
denote data; dotted red curves are fits.
\label{fig:Xib}}
\end{figure}

\begin{figure}[h]
\begin{center}
\includegraphics[width=0.49\textwidth]{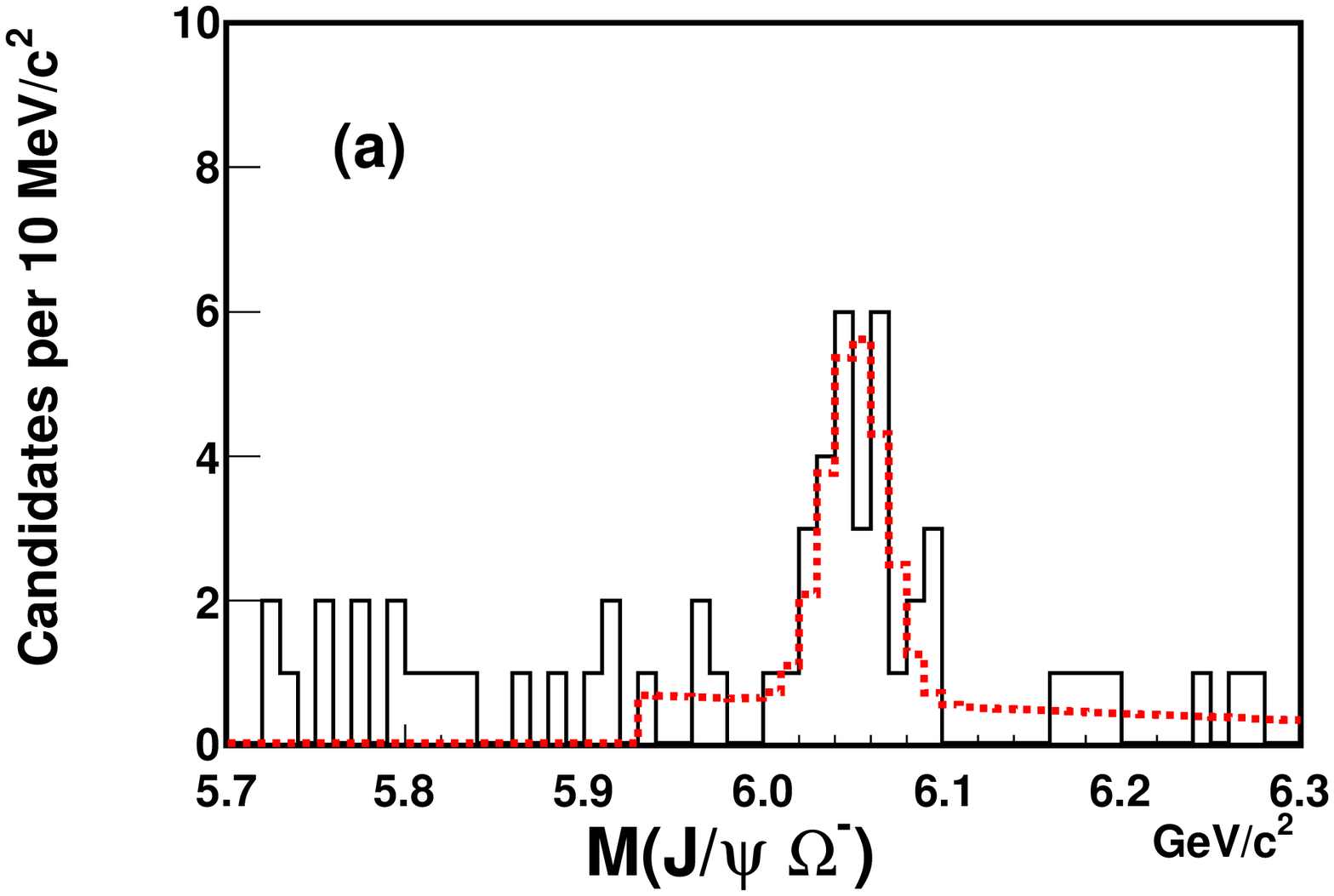}
\includegraphics[width=0.49\textwidth]{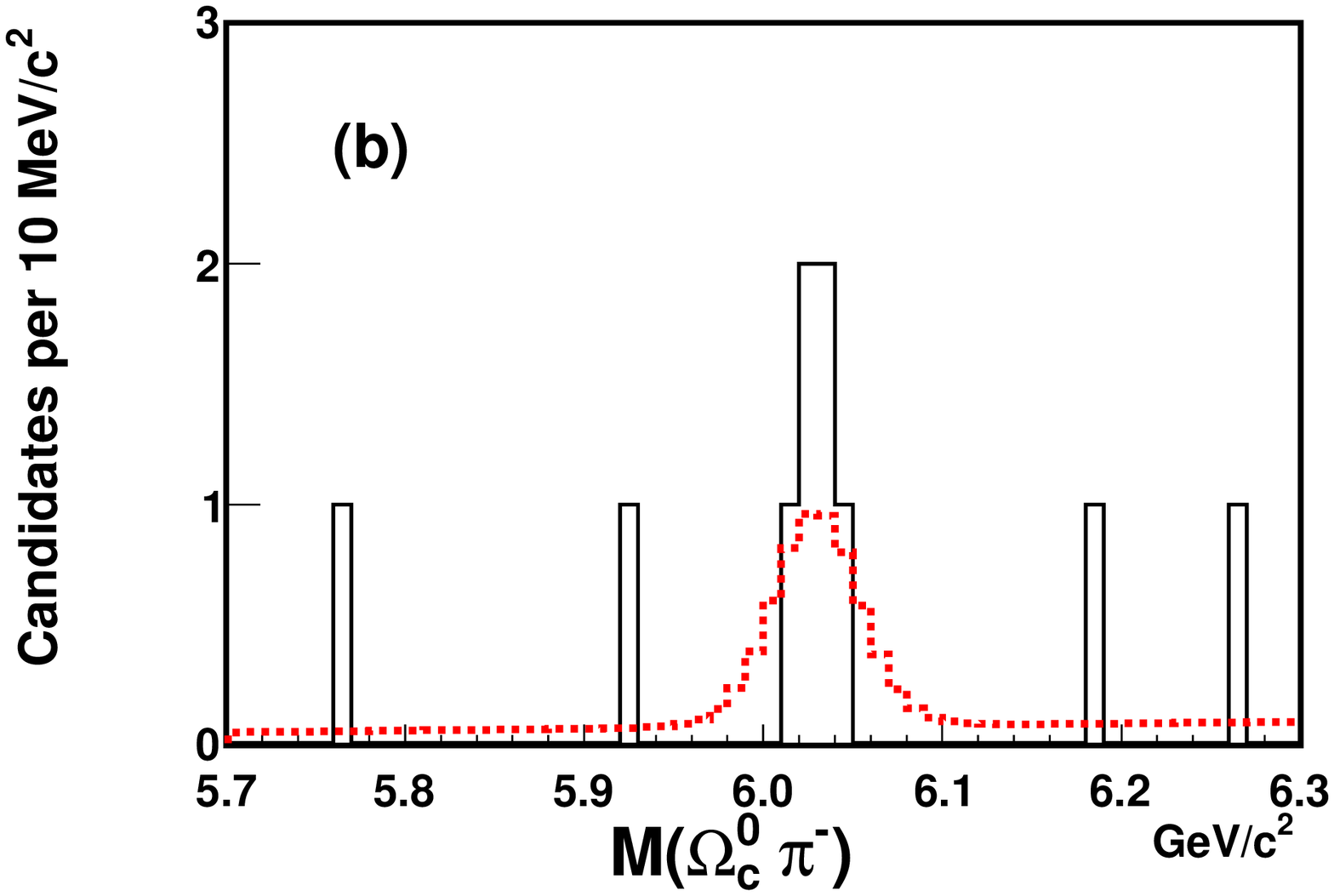}
\end{center}
\caption{Signals for $\Omega_b^- \to J/\psi \Omega^-$ (left);
$\Omega_b^- \to \Omega_c^0 \pi^-$ (right).  Curves as in Fig.\ 2.
\label{fig:Omb}}
\end{figure}

Lifetimes of singly- and doubly-strange $b$-flavored baryons have also been
measured by CDF.  The corresponding mass distributions are shown in Figs.\
\ref{fig:Xib} and \ref{fig:Omb}.

The masses obtained by CDF in these channels (as well as some charmed-strange
baryon masses) are shown in Table \ref{tab:barm}, while lifetimes are shown
in Table \ref{tab:bart}.  For reference, $\tau(B^0) = 1.519 \pm 0.005$ ps
\cite{PDG14}, while CDF and LHCb obtain $\tau(\Lambda_b)/\tau(B^0) = 1.021
\pm 0.024 \pm 0.013$ and $0.974 \pm 0.006\pm 0.004$, respectively.  A ratio
within 1\% of 1 is implied by the treatment of Ref.\ \cite{JLR}, while
Ref.\ \cite{Lenz:2014jha} finds $0.935 \pm 0.054$.

\begin{table}
\caption{Masses of charmed and $b$-flavored baryons.
\label{tab:barm}}
\begin{center}
\begin{tabular}{l c c c} \hline \hline
Baryon & Mass (MeV/$c^2$) \cite{Aaltonen:2014wfa} & PDG 2012 \cite{PDG12} &
Pred.\ \cite{Karliner:2008sv} \\ \hline
$\Xi_c^0$ & $2470.85 \pm 0.24 \pm 0.55$ & $2470.88^{+0.34}_{-0.88}$ & Input \\
$\Xi_c^+$ & $2468.00 \pm 0.18 \pm 0.51$ & $2467.8^{+0.4}_{-0.6}$ & Input \\
$\Lambda_b$ & $5620.15 \pm 0.31 \pm 0.47$ & $5619.4 \pm 0.7$ & Input (vs.\
$ \Lambda_c$) \\
$\Xi_b^-$ & $5793.4 \pm 1.8 \pm 0.7$ & $5791.1 \pm 2.2$ & $5795\pm5$\\
$\Xi_b^0$ & $5788.7 \pm 4.3 \pm 1.4$ & $5788 \pm 5$ & (Chg. avg.) \\
$\Omega_b^-$ & $6047.5 \pm 3.8 \pm 0.6$ & $(D0 \ne CDF)$ & $6052.1\pm5.6$ \\
$M(\Xi_c^0) - M(\Xi_c^+)$ & $2.85 \pm 0.30 \pm 0.04$ & $3.1^{+0.4}_{-0.5}$ &
 --\\
$M(\Xi_b^-) - M(\Xi_b^0)$ & $4.7 \pm 4.7 \pm 0.7$ & $3 \pm 6$ & $6.24\pm0.21$
 \\ \hline \hline
\end{tabular}
\end{center}
\end{table}

\begin{table}
\caption{Lifetimes of $b$-flavored baryons.
\label{tab:bart}}
\begin{center}
\begin{tabular}{l c c} \hline \hline
Baryon & CDF (ps) \cite{Aaltonen:2014wfa} & LHCb (ps) \cite{Aaij:2014owa} \\ 
\hline
$\Lambda_b$ & $1.565 \pm 0.035 \pm 0.020$ & $1.468 \pm 0.009 \pm 0.008$ \\
$\Xi_b^-$ & $1.36 \pm 0.15 \pm 0.02$ & $1.53^{+0.10}_{-0.09} \pm 0.03$ \\
$\Omega_b^-$ & $1.66^{+0.53}_{-0.40} \pm 0.02$ &
 $1.54^{+0.26}_{-0.31} \pm 0.05$ \\ \hline \hline
\end{tabular}
\end{center}
\end{table}

\begin{figure}
\begin{center}
\includegraphics[width = 0.7\textwidth]{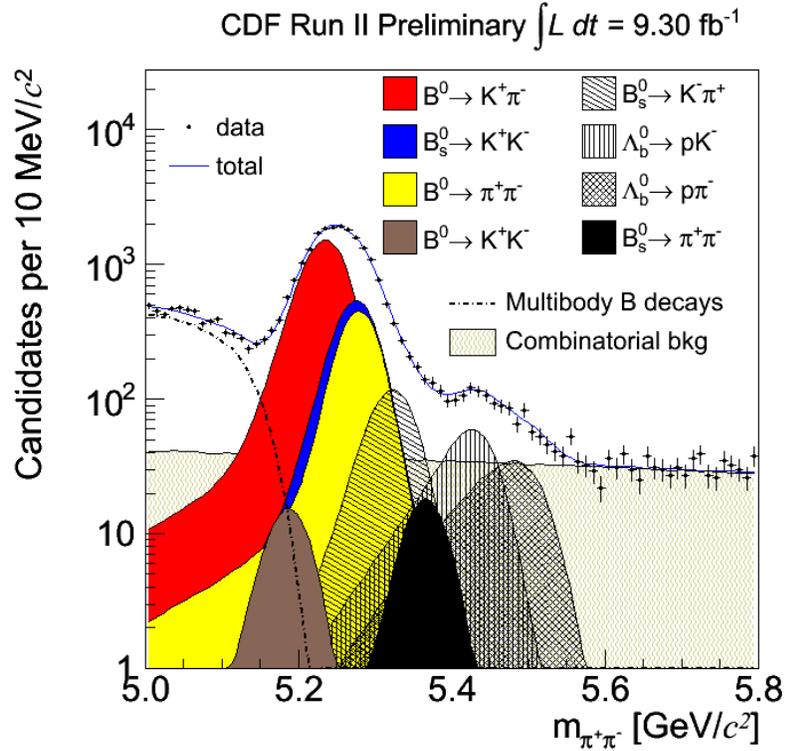}
\end{center}
\caption{CDF spectrum of $h^+h^-$ treating all $h$ as $\pi$.
\label{fig:mpipi}}
\end{figure}

\section{CP violation in charmless baryon decays}

The signal for $\Lambda_b \to h^+ h^-$ is a composite of contributions,
as shown in Fig.\ \ref{fig:mpipi}.  When plotted as if all charged particles
are pions, it leads to displaced peaks when $h \ne \pi$.  The ratio of
branching fractions ${\cal B}(\Lambda_b \to p \pi^-)/ {\cal B}(\Lambda_b \to p
K^-)$ as measured by CDF and LHCb differs from that in one
pQCD calculation (Table \ref{tab:lbrat}) which
may underestimate penguin-dominated processes such as $\Lambda_b \to p K^-$
\cite{Gronau:2013mza}.  Other decays which might be penguin-dominated are
$\Lambda_b \to \pi^\pm \Sigma^\mp$ and $\Lambda_b \to K \Xi$.

\begin{table}
\caption{Measured and predicted ratio ${\cal B}(\Lambda_b \to p \pi^-)/
{\cal B}(\Lambda_b \to p K^-)$.
\label{tab:lbrat}}
\begin{center}
\begin{tabular}{c c c} \hline \hline
Source & Reference & Value \\ \hline
CDF & \cite{Aaltonen:2008hg} & $0.66 \pm 0.14 \pm 0.08$ \\
LHCb & \cite{Aaij:2012as} & $0.86 \pm 0.08 \pm 0.05$ \\
pQCD & \cite{Lu:2009cm} & $2.6^{+2.0}_{-0.5}$ \\ \hline \hline
\end{tabular}
\end{center}
\end{table}

CDF's new results on CP violation \cite{Aaltonen:2014vra} are shown in
Table \ref{tab:ACP}.  The ratio of the first two asymmetries is as predicted
by U-spin \cite{Gronau:2000zy}. The difference between the last two asymmetries
is measured to be $(16 \pm 12)\%$, to be compared with a pQCD prediction
\cite{Lu:2009cm} of --$26\%$.

\begin{table}
\caption{CP asymmetry results from CDF \cite{Aaltonen:2014vra}.
\label{tab:ACP}}
\begin{center}
\begin{tabular}{l c c r} \hline \hline
Decay & ${\cal N}_{b \to f}$ & ${\cal N}_{\bar b \to \bar f}$ &
${\cal A}(b \to f)(\%)$ \\ \hline
$B^0 \to K^+ \pi^-$ & $5313 \pm 109$ & $6348\pm 117$ & --$8.3 \pm 1.3 \pm 0.4$
 \\
$B_s^0 \to K^- \pi^+$ & $560 \pm 51$ & $354 \pm 46$ & +$22 \pm 7 \pm 2$ \\
$\Lambda_b^0 \to p \pi^-$ & $242 \pm 24$ & $206 \pm 23$ & +$6 \pm 7 \pm 3$ \\
$\Lambda_b^0 \to p K^-$ & $271 \pm 30$ & $324 \pm 31$ & --$10 \pm 8 \pm 4$ \\
\hline \hline
\end{tabular}
\end{center}
\end{table}

\section{$B_c$ production}

\begin{figure}
\begin{center}
\includegraphics[width=0.65\textwidth]{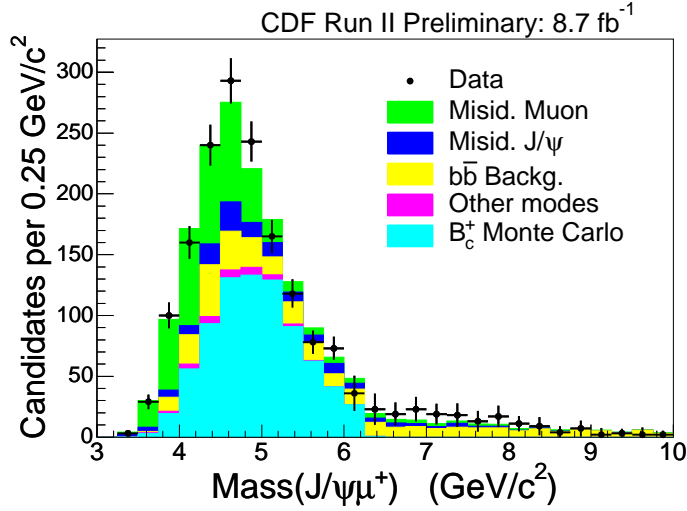}
\end{center}
\caption{$M(J/\psi \mu^+)$ spectrum including signal for $B_c \to J/\psi
\mu^+ \nu$ and background contributions.
\label{fig:bcspec}}
\end{figure}

The CDF collaboration has measured $\sigma (p \bar p \to B_c + X)
{\cal B}(B_c \to J/\psi \mu \nu)/\sigma (p \bar p \to B^+ + X){\cal B}(B^+ \to
J/\psi K^+)$ \cite{CDF11083}, with the $B^+$ decay as a normalization.
The $B_c$ decay invoves a missing neutrino, so the signal region is taken
as $4 \le M(J/\psi \mu) \le 6$ GeV/$c^2$.  Backgrounds include misidentified
$J/\psi$ plus a third muon (accounted for using $J/\psi$ sidebands), $b \bar b$
giving rise to leptons, and modes involving $\psi(2S),\tau,\ldots$.  The
fitted $M(J/\psi \mu^+)$ spectrum is shown in Fig.\ \ref{fig:bcspec}.  The
turquoise histogram labeled ``$B_c$ Monte Carlo'' is the fit to the $B_c \to
J/\psi \mu^+ \mu$ signal, with Monte Carlo shape.  The results of this fit
(and a corresponding one for $B^+ \to J/\psi K^+$) are shown in Table
\ref{tab:bcfit}.

\begin{table}[b]
\caption{Results of fits to $B_c\to J/\psi \mu^+ \mu$ and $B^+\to J/\psi K^+$
\label{tab:bcfit}}
\begin{center}
\begin{tabular}{l l} \hline \hline
Quantity & Value \\ \hline
$N(B_c^+ \to J/\psi \mu^+ \nu)$ & $739.5 \pm 39.6^{+19.8}_{-23.9}$ \\
$N(B^+ \to J/\psi K^+)$ & $14338 \pm 125$ (stat.\ only)\\
Relative efficiency & $4.093 \pm 0.038^{+0.401}_{-0.359}$ \\
$\frac{\sigma(B_c^+) * {\cal B}(B_c^+ \to J/\psi \mu^+ \nu)}
      {\sigma(B^+) * {\cal B}(B^+ \to J/\psi K^+)}$ & $0.211 \pm 0.012
^{+0.021}_{-0.020}$ \\ \hline \hline
\end{tabular}
\end{center}
\end{table}

\section{$A_{FB}(b \bar b)$ at high $m(b \bar b)$}

\begin{figure}
\begin{center}
\includegraphics[width=0.88\textwidth]{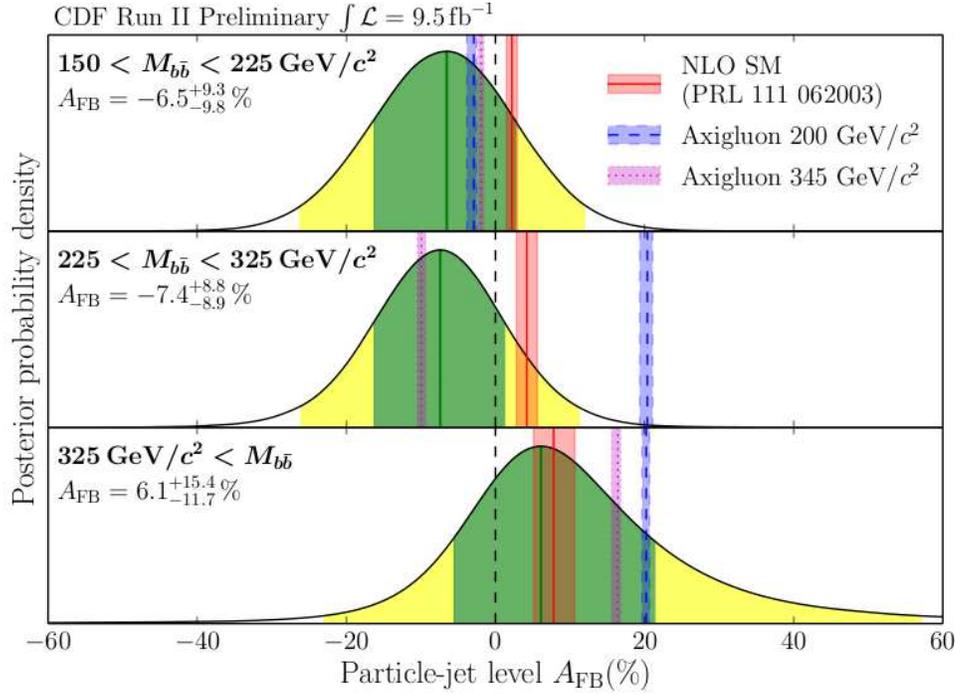}
\end{center}
\caption{CDF results on $A_{FB}(b \bar b)$ for three $M_{b \bar b}$ intervals.
Pink bands: predictions of Ref.\ \cite{Grinstein:2013mia}.
\label{fig:AFB}}
\end{figure}

The larger-than-expected $A_{FB}$ in Tevatron top quark pair production
\cite{topAFB} raises the question of whether $A_{FB}(b \bar b)$ is observable.
A proposed
``string-drag'' mechanism \cite{Rosner:2012pi} gives too small an effect
for $t \bar t$ production.  An ``axigluon'' of mass 200 GeV/$c^2$ is one
proposal \cite{Cao:2010zb} to account for the $t \bar t$
data.  The corresponding asymmetry in $q \bar q \to b \bar b$ is best probed
at high $M(b \bar b)$.  Expectations for this asymmetry in the standard model
include that of Ref.\ \cite{Grinstein:2013mia}, quoted in Fig.\ \ref{fig:AFB}
along with the CDF results \cite{CDF11092}.  The CDF values of $A_{FB}(b \bar
b)$ are consistent with the standard model and with a 345 GeV/$c^2$ axigluon
but not with a 200 GeV/$c^2$ axigluon.

\section{Excited $b$ mesons}

\begin{figure}
\includegraphics[width=0.33\textwidth]{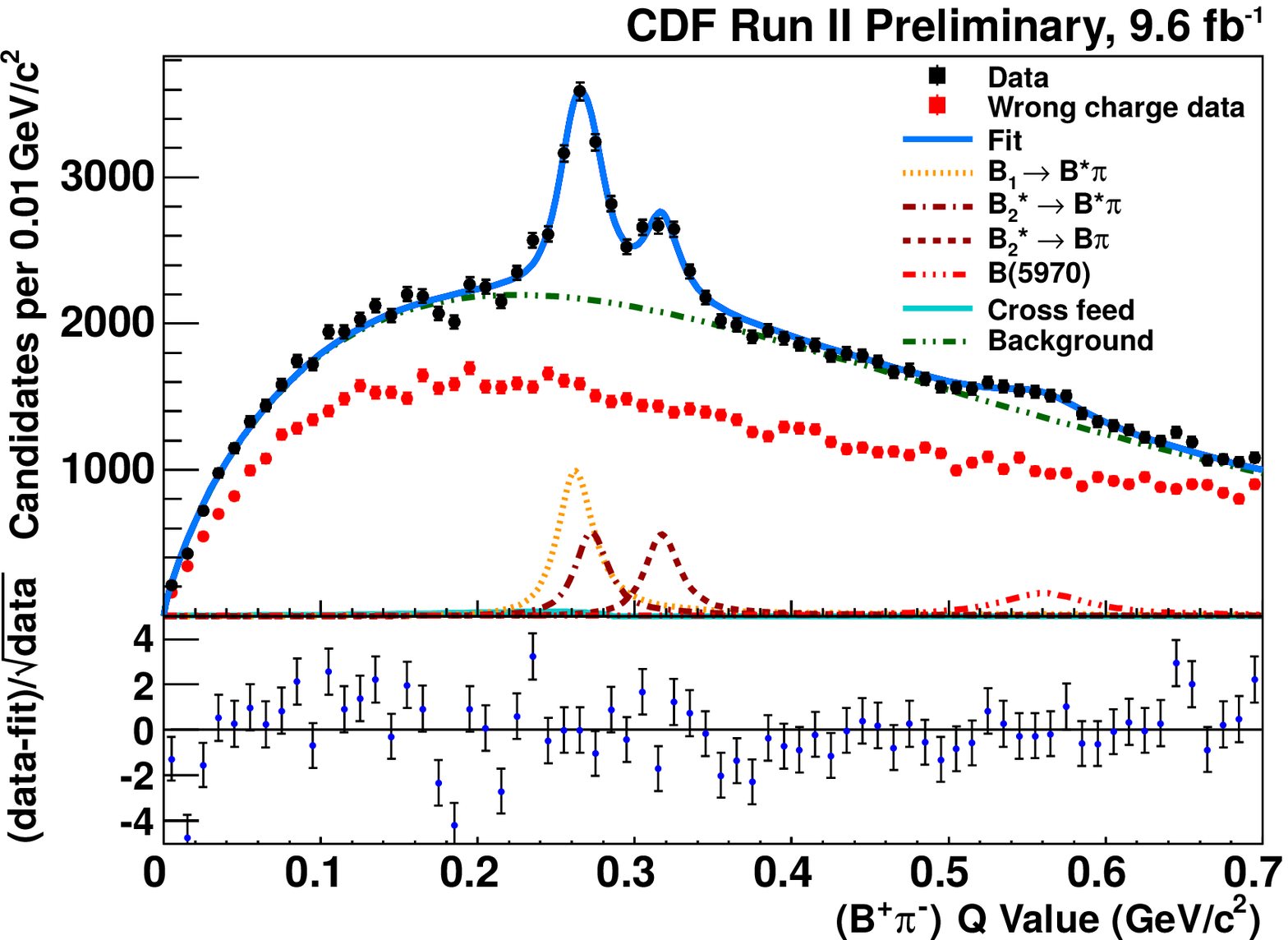}
\includegraphics[width=0.33\textwidth]{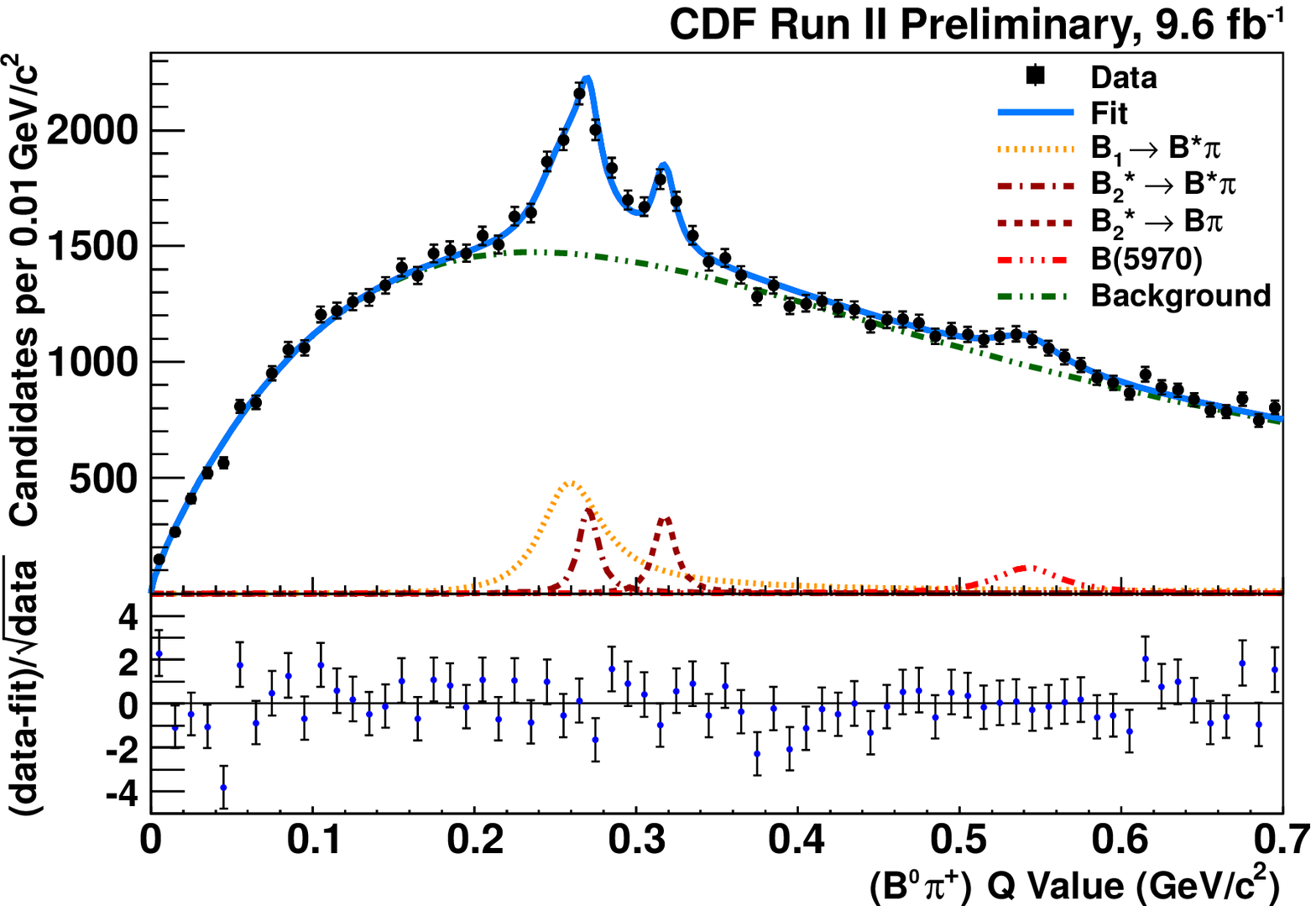}
\includegraphics[width=0.33\textwidth]{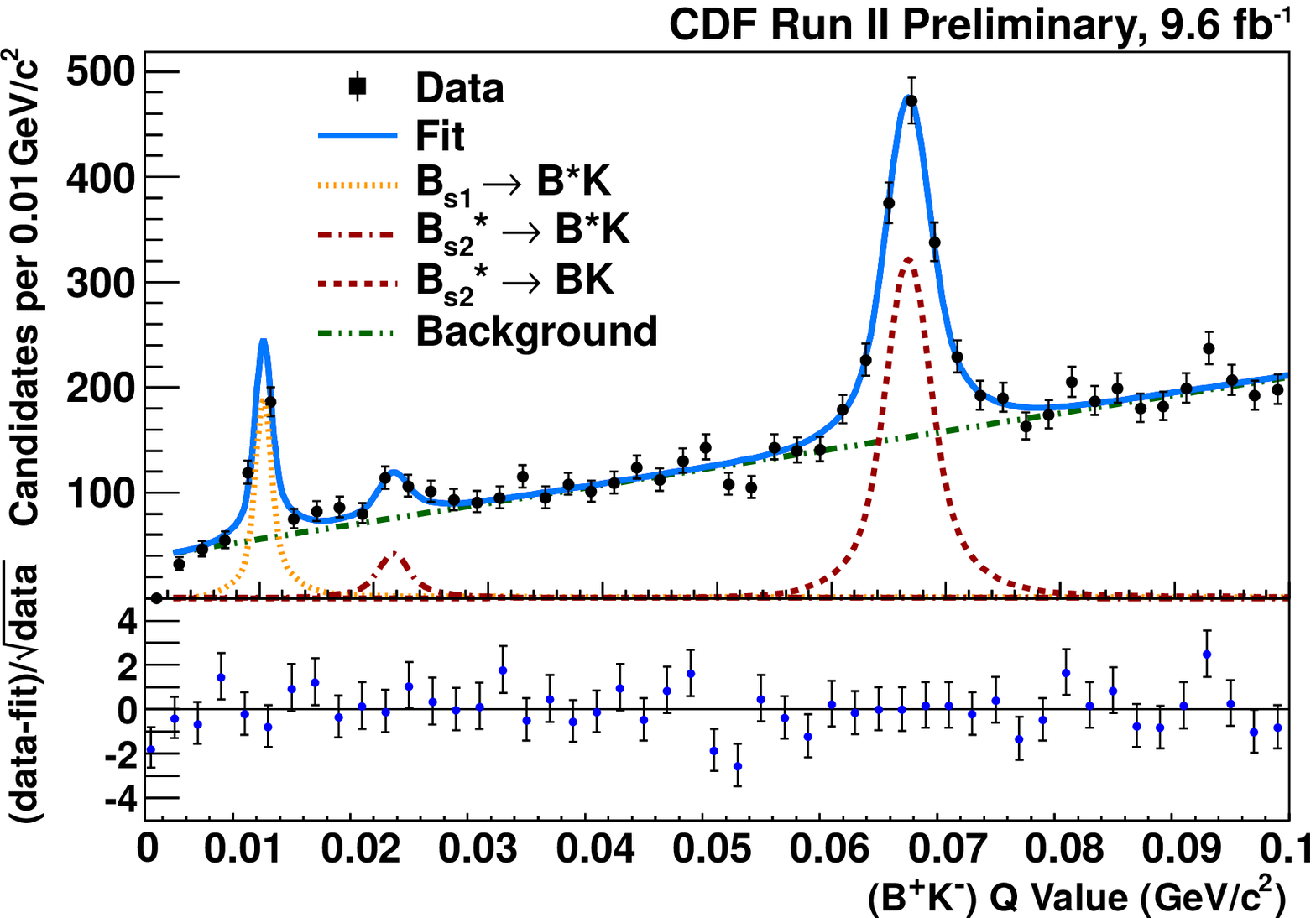}
\caption{$Q$ values for neutral $B \pi$ (left), charged $B \pi$ (middle),
and $B^+ K^-$ (right) spectra.
\label{fig:spec}}
\end{figure}

CDF has measured the masses of orbitally-excited $B$ mesons and presented
evidence for a new state decaying to $B \pi$ at 5970 MeV, as shown
in Table \ref{tab:Bstar} \cite{Aaltonen:2013atp}.  The first six decays
in Table \ref{tab:Bstar} are most likely dominated by D-waves, accounting
for the relative narrowness of these candidates for $L=1$ $\bar b d$, $\bar b
u$, and $\bar b s$ states.  The spectra on which these measurements are based
are shown in Fig.~\ref{fig:spec}.  In each of these figures, contributions to
fits are listed in order of increasing $Q$-value.

\begin{table}
\caption{Masses and widths of excited $B$ mesons reported by CDF
\cite{Aaltonen:2013atp}.  Subscripts denote total spin.
\label{tab:Bstar}}
\begin{center}
\begin{tabular}{c c c} \hline \hline
State & Mass (MeV/$c^2$) & Width (MeV/$c^2$) \\ \hline
$B_1^0$       & $5726.6 \pm 0.9^{+1.1}_{-1.2} \pm 0.4$ & $23 \pm 3 \pm 4$ \\
$B_2^{*0}$    & $5736.7 \pm 1.2^{+0.8}_{-0.9} \pm 0.2$ & $22^{+3+4}_{-2-5}$ \\
$B_1^+$       & $5727~~ \pm ~~3^{+1}_{-3}~~\pm ~~2$ & $49^{+12+2}_{-10-13}$ \\
$B_2^{*+}$    & $5736.9 \pm 1.2^{+0.3}_{-0.9} \pm 0.2$ & $11^{+4+3}_{-3-4}$ \\
$B_{s1}^0$    & $5828.3 \pm 0.1 \pm 0.2 \pm 0.4$ & $0.5 \pm 0.3 \pm 0.3$ \\
$B_{s2}^{*0}$ & $5839.7 \pm 0.1 \pm 0.1 \pm 0.2$ & $1.4 \pm 0.4 \pm 0.4$ \\
$B(5970)^0$   & $5978 \pm 5 \pm 12$ & $70^{+30}_{-20} \pm 30$ \\
$B(5970)^+$   & $5961 \pm 5 \pm 12$ & $60^{+30}_{-20} \pm 40$ \\ \hline \hline
\end{tabular}
\end{center}
\end{table}

Predicted $B_1$ masses range from 5700 to 5800 MeV/$c^2$, with $B^*_2$
5 to 20 MeV/$c^2$ higher.  Predicted $B_{s1}$ masses range from 5800
to 5900 MeV/$c^2$, with $B^*_{s2}$ 10 to 20 MeV/$c^2$ higher.  An
unquenched lattice gauge theory calculation \cite{Green:2003zza}
gives $M(B_{s1,2*}) = (5889\pm52,5901\pm52)$ MeV/$c^2$.

\section{Conclusions}

CDF has demonstrated unique capabilities for studying $b$ physics.  Baryon
masses and lifetimes agree with standard model predictions.  Charmless $b$
baryon decays show no CP violation (yet!).  The cross section times
branching ratio for $B_c$ production and decay to $J/\psi \mu \nu$ has been
measured.  The study of $A_{FB}(b \bar b)$ has ruled out an axigluon of
mass 200 GeV/$c^2$.  CDF has measured the masses and widths of orbitally
excited $B$ mesons and observed a new state at 5970 MeV decaying to $B \pi$.
All these results have exceeded expectations!

\section*{Acknowledgments}

I thank my CDF colleagues for the opportunity to present these results, and
am especially grateful to Jonathan Lewis and Patrick Lukens for valuable
advice.  This work was supported in part by the United States Department of
Energy under Grant No.\ DE-FG02-13ER41598.

\end{document}